\documentclass[ reprint,
superscriptaddress,
nobibnotes,
amsmath,amssymb,
aps,
prx,onecolumn
]{revtex4-2}

\usepackage[utf8]{inputenc}
\usepackage{amsmath}
\usepackage{amsfonts}
\usepackage{amssymb}
\usepackage{graphicx}
\usepackage{color}
\usepackage{txfonts}
\usepackage{float}
\usepackage[titletoc]{appendix}
\usepackage[colorlinks={true}]{hyperref}
\hypersetup{
    citecolor={blue},
    filecolor={blue},
    linkcolor={blue},
    urlcolor={black}
}

\usepackage{graphicx,epsfig,epstopdf}
\usepackage{amsmath} 
\usepackage{bm}
\usepackage{color}
\usepackage{subfig}
\usepackage{caption}
\usepackage{amssymb} 
\DeclareMathAlphabet{\mathcal}{OMS}{cmsy}{m}{n}
\newcommand{\be}{\begin{eqnarray}}
\newcommand{\bc}{\begin{center}}
\newcommand{\ec}{\end{center}}

\newcommand{\nn}{\nonumber}

\usepackage{centernot}
\usepackage{cancel}

\begin{document}
\title{Theoretical estimation of nonlinear response function and ultrafast optical spectroscopy of nitrogen–vacancy centers}

\author{Arunangshu Debnath}
%\email{arunangshu.debnath@desy.de}
\email{arunangshu@episteme99.com}
\affiliation{Episteme99 UG, Hamburg, Germany}

\begin{abstract}
The theoretical description of the ultrafast optical response of nitrogen-vacancy centers has traditionally been developed using the nonlinear response function formalism based on a displaced harmonic oscillator model. However, restrictive defect geometries that lead to rich phonon mode distributions can limit the validity of this model. Here, we extend this framework by incorporating state-dependent oscillator modes characterized by different frequencies. We demonstrate that, as a consequence, the nonlinear response function is modified by a straightforward correction term that can be evaluated using operator algebraic techniques. These modifications directly impact two-dimensional correlation spectroscopy, which serves as a vital characterization tool for color centers by resolving ultrafast correlated electron-vibrational dynamics.
\end{abstract}

\maketitle
\date{\today}

\section{Introduction}\label{sec:intro}
Point defects in diamond, most notably the negatively charged nitrogen-vacancy ($\mathrm{NV}^-$) color center, have emerged as preeminent platforms in both quantum sensing and information processing \cite{doherty2013nitrogen, wang2024digital, yu2021molecular, segawa2023nanoscale, fischer2025spin, dreau2011avoiding, yoon2025quantum, barry2020sensitivity, toyli2013fluorescence}.
In these systems, optical polarization and spin-state readout are assisted by spin-selective intersystem crossing pathways within the triplet manifold (${^3}A_2 \leftrightarrow {^3}E$).
Previous work has demonstrated several applications that measure coherent dynamics and estimate the phonon distributions as well as their coupling to the electronic excitation gap. A better quantitative estimation of the temporal and spectral properties of phonons affecting coherent dynamics will be essential for the scalable integration of NV centers into optical architectures. \\
Ultrafast nonlinear optical spectroscopy uses multiple laser pulses to obtain spectral and temporal insights into the dynamics of quantum emitters. These techniques are routinely applied as characterization tools in ensembles of non-interacting and interacting emitters in the condensed phase, uncovering energy and charge transfer phenomena.
The coherent optical response is interpreted in terms of interfering optical transition paths contributing to the signal. When dissipative sources affect the coherent dynamics, theoretical modeling or digital twins help relate the dissipation-modulated dynamical pathways to the optical response.
In nitrogen-vacancy centers, the two-level abstraction of optical transitions and associated phonon broadening continues to receive interest due to its relevance in the interpretation of optically detected magnetic resonance and future technological applications in sensing. Previous studies have modeled optically pumped single NV centers using kinetic equations, where rate equations have been employed to describe the dynamics. In this work, we focus on NV center ensembles.
Pioneering two-dimensional spectroscopy experiments carried out in Ref.~\cite{huxter2013vibrational} revealed the time-resolved nature of the dynamics and classified the phonon modes responsible across different frequency regimes. This work extended previous spectroscopic investigations that systematically measured linear absorption, luminescence, and dephasing \cite{drabenstedt1999low, zhang2026unraveling, ulbricht2016jahn, goldman2015phonon, su2019luminescence}.
The theoretical interpretation was facilitated by a two-level abstraction of the electronic system coupled with overdamped and multimode Brownian oscillator modes that described the dissipative dynamics. It was assumed that the ground- and excited-state potential energy surfaces are harmonic and represented by the same set of oscillators. This description, while often successful for optical transitions in simple molecules, may not remain valid in general for systems with restrictive geometries. \\
In this article, we introduce an extension of previous approaches by considering a displaced harmonic oscillator model characterized by two different sets of oscillators. We show that this has consequences for the delay-dependent spectral lineshapes measured in two-dimensional spectroscopy.\\
In the following, in Sec.~\ref{sec:theory}, we introduce the Hamiltonian models, spectral density functions, and quantify the time-dependent spectral gap function that is crucial for the evaluation of the response functions observed in two-dimensional spectroscopy. Subsequently, we develop the analytical theory of response functions, following standard derivations, and demonstrate the correction term that arises in our theory. The article concludes with a brief discussion about the measurements that may quantify these lineshape functions.
\section{Theory}\label{sec:theory}
%
%\section{prelude}
Electron-phonon interactions strictly govern the characteristic dephasing and emission characteristics of solid-state defects in color centers \cite{alkauskas2014first, gali2011ab, razinkovas2021vibrational}. In NV centers a broad phonon sideband alongside its characteristic zero-phonon line (ZPL) is visible. In the theory of nonlinear spectroscopy, these features are modeled by two electronic levels coupled to small-amplitude harmonic vibrations that represent nuclear motions (intramolecular and intermolecular vibrational modes). We consider a canonical system comprising an electronic ground state $\vert g\rangle$ and an excited state $\vert e\rangle$, coupled to an ensemble of nuclear coordinates $q$:
\begin{align}
H &= \vert g\rangle H_g(q)\langle g\vert + \vert e\rangle H_e(q)\langle e\vert,
\end{align}
where the adiabatic nuclear Hamiltonians for the respective potential energy surfaces are partitioned into the nuclear kinetic energy $T(q)$ and state-dependent potential energy terms $W_g(q)$ and $W_e(q)$: $H_g(q) = T(q) + W_g(q)$ and $H_e(q) = \hbar\omega_{eg}^0 + T(q) + W_e(q)$, with $\omega_{eg}^0$ specifying the electronic band gap between the potential energy minima. The transition dipole operator coupling the radiation field to the electronic manifolds is defined as $V = V_{ge}(q)\vert g\rangle\langle e\vert + V_{eg}(q)\vert e\rangle\langle g\vert$, where $V_{ge}(q)$ denotes an electronic transition matrix element. For the NV center, we consider the same framework, where the ground-state $^3A_2$ nuclear potential surface is modeled as a set of uncoupled harmonic oscillators, while the excited-state $^3E$ potential surface allows both the curvature (frequency) and the equilibrium nuclear position to shift. Therefore, we have
\begin{align}
H_g &= \sum_j \frac{p_j^2}{2m_{gj}} + \frac{1}{2}m_{gj}\omega_{gj}^2 q_j^2, \nonumber\\
H_e &= \hbar\omega_{eg}^\circ + \sum_j \frac{p_j^2}{2m_{ej}} + \frac{1}{2}m_{ej}\omega_{ej}^2 (q_j+d_j)^2.
\end{align}
Here, $p_j$ is the momentum operator, and $q_j$ is the position operator. The parameters $m_{gj}$ ($m_{ej}$) and $\omega_{gj}$ ($\omega_{ej}$) are the effective masses and frequencies of the ground- (excited-) state oscillators, respectively. Furthermore, $d_j$ is the spatial displacement of the equilibrium position, and $\omega_{eg}^\circ$ represents the electronic energy gap between the potential minima.
For convenience, we cast the operators into dimensionless forms defined relative to the ground-state parameters: dimensionless momentum $P_j = (\hbar\mu_{gj}\omega_{gj})^{-1/2}p_j$, dimensionless coordinate $Q_j = (\mu_{gj}\omega_{gj}/\hbar)^{1/2}q_j$, and dimensionless displacement $D_j = (\mu_{gj}\omega_{gj}/\hbar)^{1/2}d_j$. We account for the disparity in ground and excited curvatures through the dimensionless parameter $\beta_j = (\mu_{ej}\omega_{ej}/\mu_{gj}\omega_{gj})^{1/2}$. The standard linear MBO theory is recovered when taking the limit $\beta_j \rightarrow 1$. \\
% bath
The modulation of the transition energy gap, originating from electron-phonon interaction exhibits pronounced signatures of linear phonon coupling and dynamic Jahn-Teller distortion. To describe the ensuing dephasing and population dynamics, the system-bath interactions are described by a spectral density function $J(\omega) = J_{\mathrm{LFO}}(\omega) + \sum_j J_j(\omega) $; it is a combination of low-energy acoustic phonons of the bulk diamond lattice, represented as a Drude-Lorentz spectral density function and a manifold of discrete, high-frequency vibronic modes, represented as multimode Brownian oscillator function. It is given by  $\omega^2 J_{\mathrm{LFO}}(\omega) = \frac{2\lambda\tau\omega}{\tau^2 \omega^2 + 1} $ and $J_j(\omega) = \frac{2\lambda_j \omega_j^2 \omega \gamma_j}{\omega^2 \gamma_j^2 + (\omega_j^2 - \omega^2)^2}$ where $\lambda_j = S_j \hbar \omega_j $, where the mode-specific reorganization energy $\lambda_j = S_j \hbar \omega_j $ is expressed in terms of empirical Huang-Rhys factors ($S_j$) and mode frequencies ($\omega_j $). The description of the phonon modes can be generalized by adopting a generalized spectral density function \cite{debnath2013dynamics, debnath2026coherent}. \\
%
% basic algebra
In the derivation, a key step is to use algebraic manipulation to rewrite the nuclear Hamiltonians $H_g$ and $H_e$ in terms of creation ($a^\dagger$) and annihilation ($a$) operators. We adopt the operator algebra introduced in Refs.~\cite{palma1983franck, palma1992algebraic, fidler2013nonlinear, barsegov2002quantum}. We start from the standard definitions of the coordinate and momentum operators in terms of ladder operators: $q_\alpha = \sqrt{\frac{\hbar}{2\mu_\alpha\omega_\alpha}} \left( a_\alpha^\dagger + a_\alpha \right)  $ and $p_\alpha = i\sqrt{\frac{\hbar\mu_\alpha\omega_\alpha}{2}} \left( a_\alpha^\dagger - a_\alpha \right)$. The creation annihilation operators belonging to vibrational modes associated with ground and excited electronic states, represented by $a_g, a_g^\dag$ and $a_e, a_e\dag$, is given by
\begin{align}
    a_g &= \frac{\sqrt{2} \beta^2 \gamma}{1 + \beta^2} + \frac{1 - \beta^2}{1 + \beta^2} a_g^+ + \frac{2\beta}{1 + \beta^2} a_e \nn\\
    a_g^+ &= \frac{\sqrt{2} \beta^2 \gamma}{1 + \beta^2} + \frac{1 - \beta^2}{1 + \beta^2} a_g + \frac{2\beta}{1 + \beta^2} a_e^+ 
%    a_e^+ &= -\frac{\sqrt{2} \beta \gamma}{1 + \beta^2} + \frac{2\beta}{1 + \beta^2} a_g^+ - \frac{1 - \beta^2}{1 + \beta^2} a_e \nn\\
%    a_e &= -\frac{\sqrt{2} \beta \gamma}{1 + \beta^2} + \frac{2\beta}{1 + \beta^2} a_g - \frac{1 - \beta^2}{1 + \beta^2} a_e^+
\end{align}
or equivalently by
\begin{align}
%    a_g &= \frac{\sqrt{2} \beta^2 \gamma}{1 + \beta^2} + \frac{1 - \beta^2}{1 + \beta^2} a_g^+ + \frac{2\beta}{1 + \beta^2} a_e \nn\\
%    a_g^+ &= \frac{\sqrt{2} \beta^2 \gamma}{1 + \beta^2} + \frac{1 - \beta^2}{1 + \beta^2} a_g + \frac{2\beta}{1 + \beta^2} a_e^+ \nn\\
    a_e^+ &= -\frac{\sqrt{2} \beta \gamma}{1 + \beta^2} + \frac{2\beta}{1 + \beta^2} a_g^+ - \frac{1 - \beta^2}{1 + \beta^2} a_e \nn\\
    a_e &= -\frac{\sqrt{2} \beta \gamma}{1 + \beta^2} + \frac{2\beta}{1 + \beta^2} a_g - \frac{1 - \beta^2}{1 + \beta^2} a_e^+
\end{align}
which can be recast as transformation reminiscent of Bogoliubov transformation
\begin{align}
   \begin{pmatrix} a_g \\ a_g^\dagger \end{pmatrix} = \begin{pmatrix} u & -v \\ -v & u \end{pmatrix} \begin{pmatrix} a_e \\ a_e^\dagger \end{pmatrix} + \begin{pmatrix} \delta \\ \delta 
   \end{pmatrix}
\end{align}
where $ u = \frac{\beta^2 + 1}{2\beta} $, $v = \frac{\beta^2 - 1}{2\beta}$ and $ \delta = \frac{\gamma}{\sqrt{2}}$. We also define $\beta = \left(\frac{m_E w_E}{m_G w_G}\right)^{1/2}d$, and $\gamma = \left(\frac{m_G w_G}{\hbar}\right)^{1/2} $.  \\
It leads to
\begin{align}
    H_g &= \frac{p_{gj}^2}{2\mu_{gj}} + \frac{1}{2}\mu_{gj}\omega_{gj}^2 q_{gj}^2 \nn\\
    &= \frac{\hbar\omega_{gj}}{4} \left[ -(a_{gj}^\dagger - a_{gj})^2 + (a_{gj}^\dagger + a_{gj})^2 \right] \nn\\
    &= \frac{\hbar\omega_{gj}}{2} \left( a_{gj}^\dagger a_{gj} + a_{gj} a_{gj}^\dagger \right)
\end{align}
Using the bosonic commutation relation $[a_{gj}, a_{gj}^\dagger] = 1 \implies a_{gj} a_{gj}^\dagger = a_{gj}^\dagger a_{gj} + 1$, we get $H_g = \sum_j \hbar\omega_{gj} \left( a_{gj}^\dagger a_{gj} + \frac{1}{2} \right)$. Similarly, the excited-state Hamiltonian is given by
\begin{align}
    H_e &= \hbar\omega_{eg}^\circ + \sum_j \left[ \frac{p_{j}^2}{2\mu_{ej}} + \frac{1}{2}\mu_{ej}\omega_{ej}^2(q_j + d_j)^2 \right] \nn\\
    &=  \hbar\omega_{eg}^\circ + \sum_j \left( \frac{p_{ej}^2}{2\mu_{ej}} + \frac{1}{2}\mu_{ej}\omega_{ej}^2 q_{ej}^2 \right) \nn\\
    &=  \hbar\omega_{eg}^\circ + \sum_j \hbar\omega_{ej} \left( a_{ej}^\dagger a_{ej} + \frac{1}{2} \right)
\end{align}
where we have used $q_{ej} = q_{gj} + d_j$ and $p_{ej} = p_{gj}$. For the computation, however, it is preferable to express $H_e$ in terms of the ground-state operators. Following straightforward algebraic manipulation, we arrive at
\begin{align}
    H_e &= \hbar\omega_{eg}^\circ + \sum_j \hbar\omega_{ej} \bigg[ \frac{\beta_j^4 - 1}{4\beta_j^2} \left( (a_{gj}^\dagger)^2 + a_{gj}^2 \right) \nn\\&+ \frac{1 + \beta_j^4}{2\beta_j^2} \left( a_{gj}^\dagger a_{gj} + \frac{1}{2} \right) + \frac{D_j \beta_j^2}{\sqrt{2}} (a_{gj}^\dagger + a_{gj}) + \frac{D_j^2 \beta_j^2}{2} \bigg]
\end{align}
When $\beta_j = 1$ and $\omega_{ej} = \omega_{gj}$, the squeezing/two-phonon term $(a_{gj}^\dagger)^2 + a_{gj}^2$ vanishes, reducing directly to the standard displaced multimode Brownian oscillator Hamiltonian.\\
The theoretical formalism developed by Mukamel and coworkers \cite{mukamel1995principles, hamm2011concepts, cho2009two} have been foundation to two-dimensional spectroscopy and evaluation of nonlinear response function. Following these derivations, the collective gap coordinate is defined as
\begin{align}
    U & = H_e - H_g - \hbar\omega_{eg}
\end{align}
where $\omega_{eg}$ represents a reference frequency chosen to effectively truncate the moment expansions.
This quantity is related to the thermally averaged electronic energy gap (in the condon approximation) and is given by 
\begin{align}
   \omega_{eg} &\equiv \langle(H_e - H_g)\rho_g\rangle \nn\\
   &= \omega_{eg}^0 + \frac{1}{\hbar}\langle(W_e - W_g)\rho_g\rangle  
\end{align} 
The thermalized energy gap is given by $\hbar\omega_{eg} = \langle (H_e - H_g) \rho_g \rangle$, where the ground-state thermal density matrix is $\rho_g = e^{-H_g / k_b T} / \langle e^{-H_g / k_b T} \rangle$
The centered energy gap operator $U$ representing thermal fluctuation %will have the moments satisfy $M_0 = 1$, $M_1 = 0$, and $M_n = \langle U^n \rangle$ which 
helps the analytical derivation. We are interested in $U(\tau)$ which is given by 
\begin{align}
   U(\tau) &\equiv \exp\left(\frac{i}{\hbar}H_g\tau\right)U\exp\left(-\frac{i}{\hbar}H_g\tau\right) 
\end{align}
We can use these relations to express transition operators in terms of time-ordered exponentials:
\begin{align}
    V_{ge}(\tau) &= \exp(-i\omega_{eg}\tau)\exp_+\left[-i\int_0^\tau d\tau'\, U(\tau')\right] \nn\\
    V_{eg}(\tau) &= \exp(i\omega_{eg}\tau)\exp_-\left[i\int_0^\tau d\tau'\, U(\tau')\right] 
\end{align}
which leads to
%Substituting into the definition of $F(\tau_1,\tau_2,\tau_3,\tau_4)$ yields:
\begin{align}
    & F(\tau_1,\tau_2,\tau_3,\tau_4) = \exp[-i\omega_{eg}(\tau_1 - \tau_2 + \tau_3 - \tau_4)] \nn\\ &
    \times \left\langle \exp_+\left[-i\int_0^{\tau_1} d\tau_1'\, U(\tau_1')\right] \exp_-\left[i\int_0^{\tau_2} d\tau_2'\, U(\tau_2')\right] \right. \nn\\ &
   \times \left. \exp_+\left[-i\int_0^{\tau_3} d\tau_3'\, U(\tau_3')\right] \exp_-\left[i\int_0^{\tau_4} d\tau_4'\, U(\tau_4')\right] \right\rangle
\end{align}
In the traditional displaced harmonic oscillator model, expanding the time-ordered (and anti-time-ordered) exponential via cumulants leads to all higher-order cumulants factorizing into lower-order ones. The resulting algebraic steps have been formalized extensively in the literature.
We define
\begin{align}
g(\tau_1) &\equiv \int_0^{\tau_1} \mathrm{d}\tau \int_0^\tau \mathrm{d}\tau', \langle U(\tau)U(\tau') \rangle, \\
h(\tau_1, \tau_2) &\equiv \int_0^{\tau_1} \mathrm{d}\tau \int_0^{\tau_2} \mathrm{d}\tau', \langle U(\tau)U(\tau') \rangle.
\end{align}
Using these definitions, the time-dependent lineshape functions appearing under the integral can be evaluated for the $\beta=1$ case using a cumulant expansion. The derivation yields
\begin{align}
& F(\tau_1, \tau_2, \tau_3, \tau_4) = \exp\bigg[-i\bar{\omega}_{eg}(\tau_1 - \tau_2 + \tau_3 - \tau_4)\bigg] \nonumber \\
&\quad \times \bigl[1 - g^*(\tau_1) - g^*(\tau_2) - g^*(\tau_3) - g^*(\tau_4) \nonumber \\
&\quad\qquad + h(\tau_1, \tau_2) - h(\tau_1, \tau_3) + h(\tau_2, \tau_3) \nonumber \\
&\quad\qquad + h(\tau_1, \tau_4) - h(\tau_2, \tau_4) + h(\tau_3, \tau_4) + \cdots\bigr]
\end{align}
which can be reduced to
\begin{align}
& F(\tau_1, \tau_2, \tau_3, \tau_4) = \exp\bigg[-i\bar{\omega}_{eg}(\tau_1 - \tau_2 + \tau_3 - \tau_4)\bigg] \nonumber \\
&\quad \times \bigl[1 - g(\tau_1 - \tau_2) + g(\tau_1 - \tau_3) - g(\tau_2 - \tau_3) \nonumber \\
&\quad\qquad - g(\tau_1 - \tau_4) + g(\tau_2 - \tau_4) - g(\tau_3 - \tau_4) + \cdots\bigr]
\end{align}
using the relation $h(\tau_1, \tau_2) = g(\tau_1) - g(\tau_1 - \tau_2) + g(-\tau_2)$. 
The key quantities of interest in the case of displaced harmonic oscillators with differing effective harmonic properties (i.e., when $\beta \neq 1$) are the two-point and three-point correlation functions, denoted by $C^{(2)}(t_1)$ and $C^{(3)}(t_1, t_2)$, respectively. They are given by
\begin{align}
    C^{(2)}(t_1) &= \frac{1}{\hbar^2}\langle U(t_1)U(0)\rho_g\rangle \nn\\
    C^{(3)}(t_1, t_2) &= \frac{1}{\hbar^3}\langle U(t_2)U(t_1)U(0)\rho_g\rangle
\end{align}
which leads to
\begin{align}
g(t) &=\int_0^t d\tau_2\int_0^{\tau_2} d\tau_1 C^{(2)}(\tau_2-\tau_1) \nn\\
%\end{align}
%\begin{align}
g_c(t) &= \int_0^t d\tau_3\int_0^{\tau_3} d\tau_2\int_0^{\tau_2} d\tau_1 C^{(3)}(\tau_2-\tau_1, \tau_3-\tau_1)
\end{align}
We also define 
%case we have three sets of definitions 
\begin{align}
h_1(t_1, t_2) &= \frac{1}{\hbar^3} \int_0^{t_1}\mathrm{d}\tau_1 \int_0^{\tau_1}\mathrm{d}\tau_2 \int_0^{t_2}\mathrm{d}\tau_3\, \langle U(\tau_1) U(\tau_2) U(\tau_3) \rangle_g \\
h_2(t_1, t_2) &= \frac{1}{\hbar^3} \int_0^{t_1}\mathrm{d}\tau_1 \int_0^{\tau_1}\mathrm{d}\tau_2 \int_0^{t_2}\mathrm{d}\tau_3\, \langle U(\tau_2) U(\tau_1) U(\tau_3) \rangle_g \\
h_3(t_1, t_2, t_3) &= \frac{1}{\hbar^3} \int_0^{t_1}\mathrm{d}\tau_1 \int_0^{t_2}\mathrm{d}\tau_2 \int_0^{t_3}\mathrm{d}\tau_3\, \langle U(\tau_1) U(\tau_2) U(\tau_3) \rangle_g
\end{align}
The resonant third-order nonlinear response in the case of displaced harmonic oscillators with differing effective harmonic properties is determined by
\begin{align}
   & \langle V_{ge}(t_1) V_{eg}(t_2) V_{ge}(t_3) V_{eg}(t_4)\rangle_g  = \langle V_{ge} V_{eg} V_{ge} V_{eg}\rangle \nn\\&
   \exp[-i\omega_{eg}(t_1-t_2+t_3-t_4)] 
%   L^{(2)}(t_1,t_2,t_3,t_4) F_c(t_1,t_2,t_3,t_4) 
     F(t_1,t_2,t_3,t_4) F_c(t_1,t_2,t_3,t_4) 
\end{align}
which depends explicitly on both the second and third-order cumulant contributions, via the normal and correction terms, $ F(t_1,t_2,t_3,t_4)  $ and $F_c(t_1,t_2,t_3,t_4)$, respectively. 
The normal term is given by
\begin{align}
 &F(t_1, t_2, t_3, t_4) = \exp\bigl[-g(t_1 - t_2) + g(t_1 - t_3)  \nonumber \\
&\quad - g(t_2 - t_3) - g(t_1 - t_4) + g(t_2 - t_4) - g(t_3 - t_4)\bigr]
\end{align}
and the correction term is given by
\begin{align}
& F_c(t_1, t_2, t_3, t_4) = \exp\Bigl[ i g_c(t_1) - i {g_c}^*(t_2) + i g_c(t_3) - i {g_c}^*(t_4) \nonumber \\
&\quad - i h_3(t_1, t_2, t_3) + i h_3(t_1, t_2, t_4) - i h_3(t_1, t_3, t_4) + i h_3(t_2, t_3, t_4) \nonumber \\
&\quad + i {h_1}^*(t_2, t_1) + i {h_2}^*(t_3, t_1) + i {h_1}^*(t_4, t_1) - i h_1(t_1, t_2) \nonumber \\
&\quad + i h_1(t_1, t_3) - i h_1(t_1, t_4) - i {h_2}^*(t_3, t_2) - i {h_1}^*(t_4, t_2) \nonumber \\
&\quad + i h_2(t_2, t_3) - i h_2(t_2, t_4) + i {h_1}^*(t_4, t_3) - i h_1(t_3, t_4) \Bigr]
\end{align}
In traditional theoretical models, the ground- and excited-state potential energy surfaces are assumed to have the same curvature, usually taken to be harmonic. A minimal extension to this model is to introduce electronic states with differing curvatures, characterized by coupling to distinct sets of harmonic oscillator modes. Its consequence on the nonlinear response function is reflected in a correction term.\\
%
%
%%%%%%%%%%%%%%%%%%%%%%
\begin{figure}[ht!]
%\centering
 \includegraphics[width=.48\textwidth]{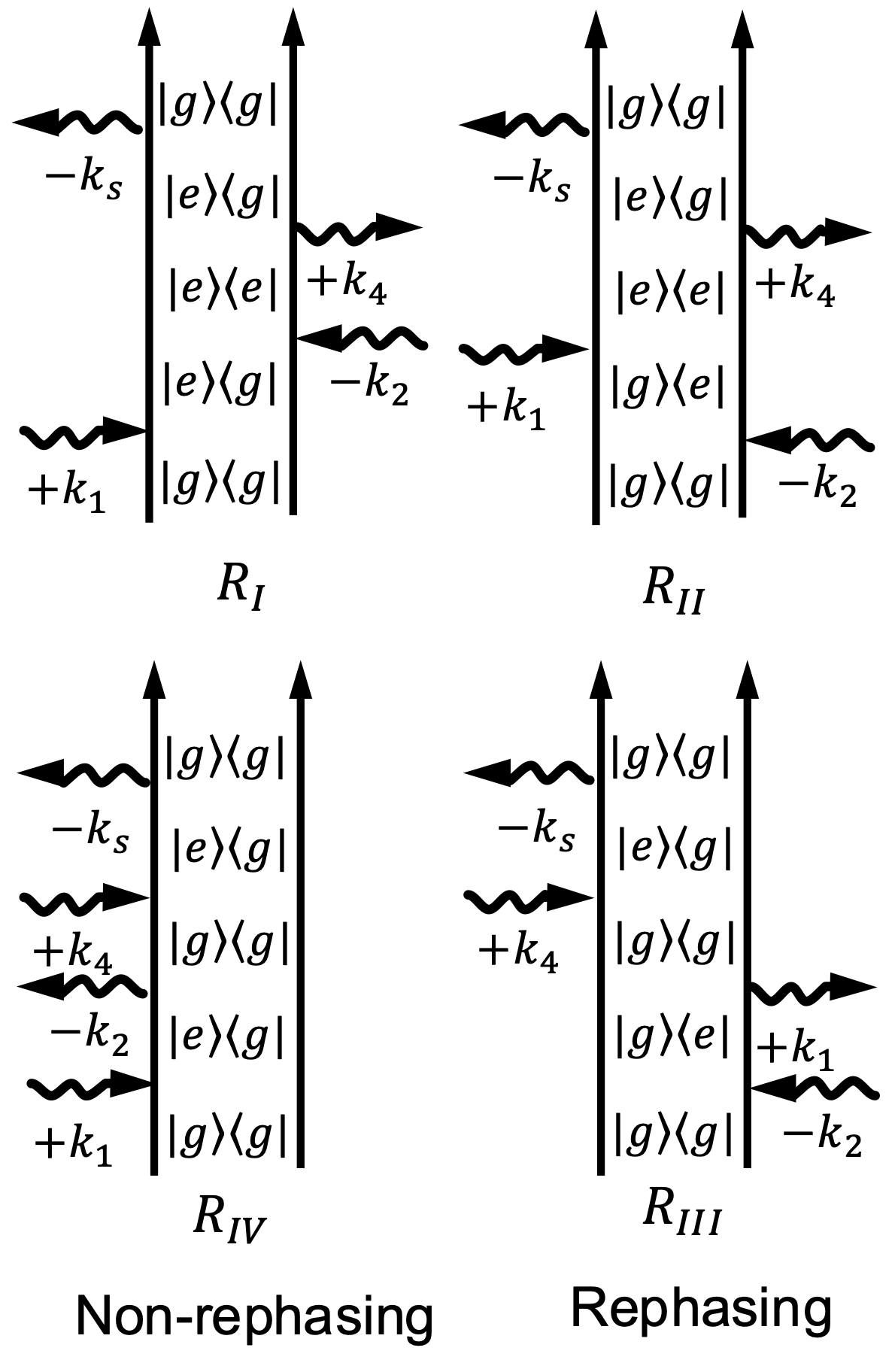}
 \caption{Liouville space diagrams representing four of the pathways that contribute to the signal. During the intra-pulse delays, the system undergoes dissipative evolution. The lineshape functions are dependent on the delays.}
  \label{fig:feyndiag}
\end{figure}
%%%%%%%%%%%%%%%%%%%%%%%%
%
The nonlinear response function without the correction terms are given by 
\begin{align}
S^{(3)}(t_3, t_2, t_1) &= \bigg(\frac{i}{\hbar}\bigg)^3 \theta(t_1)\theta(t_2)\theta(t_3) \sum_{\alpha=1}^4 \bigg[ R_\alpha(t_3, t_2, t_1) - R_\alpha^*(t_3, t_2, t_1) \bigg] 
\end{align}
where the pathways are defined as
\begin{align}
R_1(t_3, t_2, t_1) &= \mu_{ge}^2 \mu_{eg}^2\exp\bigg(-i\omega_{eg}t_1 - i\omega_{eg}t_3\bigg) \nn\\&
\exp\bigg[-g^*(t_3) - g(t_1) 
- f_+(t_3, t_2, t_1)\bigg] \\
R_2(t_3, t_2, t_1) &= \mu_{ge}^2 \mu_{eg}^2\exp\bigg(i\omega_{eg}t_1 - 
i\omega_{eg}t_3\bigg) \nn\\& \exp\bigg[-g^*(t_3) - g^*(t_1) 
+ f_+^*(t_3, t_2, t_1)\bigg] \\
R_3(t_3, t_2, t_1) &= \mu_{ge}^2 \mu_{eg}^2\exp\bigg(i\omega_{eg}t_1 - 
i\omega_{eg}t_3\bigg) \nn\\& \exp\bigg[-g(t_3) - g^*(t_1)
+ f_-^*(t_3, t_2, t_1)\bigg] \\
R_4(t_3, t_2, t_1) &= \mu_{ge}^2 \mu_{eg}^2\exp\bigg(-i\omega_{eg}t_1 - i\omega_{eg}t_3\bigg) \nn\\& \exp\bigg[-g(t_3) - g(t_1) 
- f_-(t_3, t_2, t_1)\bigg]
\end{align}
where
\begin{align}
f_-(t_3, t_2, t_1) &\equiv g(t_2) - g(t_2 + t_3) - g(t_1 + t_2) + g(t_1 + t_2 + t_3) \\
f_+(t_3, t_2, t_1) &\equiv g^*(t_2) - g^*(t_2 + t_3) - g(t_1 + t_2) + g(t_1 + t_2 + t_3)
\end{align}
Therefore, within this model, the time-delay dependent lineshape functions augment the coherent oscillations and play key role ind determining the response.
\section{Conclusion}\label{sec:conclusion}
In this work, we introduced a theoretical extension of the lineshape model used in two-dimensional spectroscopy to facilitate a more accurate interpretation of measurements carried out on NV center ensembles. These measurements can offer novel insights into the optical and vibronic dynamics of the center at sub-picosecond timescales and low temperatures. Previous studies have shown that the optical absorption profile is dominated by the zero-phonon line (ZPL) at $638\text{ nm}$ alongside a pronounced, asymmetric phonon sideband that extends over $300\text{ meV}$ toward higher energies.
Conventional steady-state or single-frequency spectroscopic probes cannot resolve the microscopic energy flow between electronic and vibrational modes or quantify homogeneous dephasing mechanisms.
In ensembles, locally fluctuating environments obscure the intrinsic homogeneous transition linewidths. Multidimensional coherent spectroscopy techniques provide a way to separate homogeneous dephasing from static disorder and quantify ultrafast spectral diffusion as well as non-Markovian bath fluctuations. This approach holds promise as an essential characterization tool for color centers in general and the NV center specifically.\\
We have limited our description to weak-field, multi-pulse spectroscopy. The classical driving field used here can be replaced by entangled photon sources \cite{debnath2026entangled, debnath2026photon, debnath2022entangled}, which can provide high-resolution, time-resolved dynamics while scaling linearly with the intensity of the incident field. The description of the phonon bath can be extended by adopting a generalized spectral density function \cite{debnath2013dynamics, debnath2026coherent}. Excited-state spectroscopy was previously demonstrated in Ref.~\cite{fuchs2008excited}. Finally, we note that similar studies can be extended to excited-state electronic transitions and to configurations in the presence of finite magnetic fields, the latter being particularly well-suited for sensing applications \cite{barry2020sensitivity}.

\begin{acknowledgments}
A.D. acknowledges funding from IFB (Hamburgische Investitions- und Förderbank), Hamburg, and infrastructure facilities made available via Start-up Labs Bahrenfeld, Hamburg, Germany.\end{acknowledgments}

\bibliography{apssamp}
\end{document}